\documentstyle[preprint,eqsecnum,aps]{revtex}
\begin{document}
\draft

\title{Phase Diagram of the Two-Dimensional Anderson-Hubbard Model}

\author{Jaichul Yi, Lizeng Zhang and Geoff S. Canright}

\address{
Department of Physics and Astronomy, University of Tennessee\\
Knoxville, TN 37996-1200  \\ and \\
Solid State Division, Oak Ridge National Laboratory \\
Oak Ridge, TN 37831}

\maketitle
\begin{abstract}
We study the ground state of the two-dimensional Anderson-Hubbard
model using a quantum real space renormalization group method. We
obtain the phase diagram near half filling. The system is always
insulating with disorder. At half filling, the system undergoes
a transition from a gapless (Anderson) insulator to an incompressible
(Mott-Hubbard) insulator as the interaction $U$ reaches
a critical value $U_c$. Away from half filling, the insulating
phase is always gapless and is found to be controlled by the
Anderson fixed point at half filling. This result is similar to
that obtained in the corresponding one dimensional system
and suggests strongly the importance of the electron-electron
correlation in this gapless insulating phase.

\end{abstract}

\pacs{}

\narrowtext
\section{Introduction}

When disorder is introduced into a physical system, it will result
in the localization of some single particle states \cite{anderson1}.
For a noninteracting electron system, if the states at the
fermi surface are localized, the system is a so-called `Anderson
insulator'. In reality, the interaction between electrons always
exists, and  such a single-particle picture may not apply.
The understanding of the interplay between disorder
and interaction has been an important issue in condensed matter
physics\cite{lee}. This issue has become more interesting since
the discovery of the high temperature superconductors
\cite{muller}. On the one hand, it is commonly believed that
these materials are strongly two dimensional (2D) in character,
and that electron-electron
correlations are important and are responsible for many of their
unusual physical properties; on the other hand, it is also clear
that disorder, which  manifests itself as (e.g.,) oxygen
vacancies, is inevitably present in these materials. It is
thus of interest to investigate the effect of disorder in highly
correlated systems. In this paper, we pursue such a study
using a real space renormalization group (RG) approach.
The model we consider is the two dimensional
Anderson-Hubbard model defined by the Hamiltonian
\begin{equation}\label{hah}
H=\sum_{\scriptstyle<i,j>\atop\scriptstyle\sigma}
  ( t_{ij}c^{\dagger}_{i\sigma}c_{j\sigma} + H.c. )
   + \sum_{i\sigma}(W_{i}-\mu)n_{i\sigma}
   +U\sum_{i}n_{i\uparrow}n_{i\downarrow}.
\end{equation}
Here $c^{\dagger}_{i\sigma}(c_{i\sigma})$ is the creation
(annihilation) operator for a spin-$\sigma$ electron on site $i$,
$t_{ij}$ is the nearest-neighbor intersite hopping energy and
$U$ ($>0$) gives the on-site Coulomb repulsion energy. The chemical
potential is given by $\mu$, and $W_i$ is a random site potential
which has an independent gaussian distribution with zero mean and
width $W$, i.e., ${\overline W_i}=0$ and
${\overline {W_iW_j}}=W^2\delta_{ij}$ (where the overbar
indicates random average). We shall consider only the
square lattice case.  Without interaction ($U=0$) but
with disorder, this is the Anderson model \cite{anderson1}
of localization which has been the prototype for studying
the effect of disorder in electron systems.
With interaction but without disorder ($W_{i} \equiv 0$), this
is the Hubbard model \cite{hubbard}, which is believed to be
one of the simplest theoretical models which possesses the
essential physics of correlation \cite{anderson2}, and which
has been the focus of intense theoretical investigation since
the discovery of high $T_c$ superconductors. Thus, the
Anderson-Hubbard model is a natural
starting point for the investigation of the combined effects
of disorder and interaction in electron systems.
Throughout the paper our discussion will be restricted to
ground state properties.

For the Anderson model in 2D, the consensus is that all of
its eigenstates are localized, and hence that it describes a
gapless insulating state \cite{lee}. For the Hubbard model
at half filling, on the other hand, it is commonly accepted
that the Coulomb repulsion $U$ gives rise to insulating behavior,
with long-ranged antiferromagnetic ordering in the ground state.
Thus, in contrast to the usual `band insulator' where the
insulating phase is due to the filling of electron bands in
the solid, and different from the Anderson insulator where the
vanishing conductivity is caused by the localization of the
single particle states at the fermi surface, the insulating state
in the Hubbard model at half filling is a result of
electron-electron correlations, hence a `correlated
insulator'. Away from but close to half filling, the Hubbard
model describes a highly correlated system
whose exact properties we still know little about despite
intense studies during the past few years. Candidates for
the possible ground states can be, e.g., phase separation
\cite{emery}, a highly correlated metal \cite{review}, or
a superconductor \cite{anderson2,review}.
When both interaction and disorder are considered, one
expects that disorder breaks the translation and other
lattice symmetries and possibly weakens effect of the
correlations; on the other hand, strong correlations may
render the standard single particle picture of Anderson
localization meaningless.  As a first step towards
understanding this complicated issue, we wish to identify
the phase diagram of the 2D Anderson-Hubbard model.

Previously, the Landau Fermi-liquid idea has been employed to
describe systems of weakly interacting electrons with (weak)
disorder \cite{lee}. The validity of approaches along this line
is questionable in the present situation because the
noninteracting system is non-metallic, and because the effect of
the interaction is presumably nonperturbative. The real space
RG approach \cite{hirsch}--\cite{singh}, on the other hand, is a
nonperturbative method which allows one to treat disorder
and interaction of any strength on the same footing.
It has however the disadvantage of being an uncontrolled
approximation, so that its implementation and interpretation
should be taken with extra caution.

The real space RG scheme adopted in this paper is a
generalization of the works of Hirsch \cite{hirsch} and Ma
\cite{ma}. This method allows one to study the compressibility
of the system by investigating the renormalization of chemical
potential and the corresponding flow of density.  This RG scheme
has been previously employed to study the $U = \infty$
Anderson-Hubbard model for spinless bosons \cite{lizma,lizwa}.
While the quantitative results, such as the critical exponents
of the superfluid--Bose-glass phase transition, are still the
subject of some controversy \cite{dirtyB}, this method does
provide the correct qualitative physical picture. For instance,
it shows that the superfluid phase is unstable against any
amount of disorder in the 1D $U=\infty$ Anderson-Hubbard model,
in agreements with the exact result \cite{lizma2}; in 2D and 3D,
it shows a transition from the superfluid phase to a disordered
(Bose-glass) phase at some critical amount of disorder, as
indicated by other theoretical approaches \cite{dirtyB}.  Thus
we have reasons to believe that the real space RG approach can
also give us useful information concerning the fermion
Anderson-Hubbard model (\ref{hah}).  In addition, the validity
of the real space RG scheme for the present case may be tested
in the noninteracting system; for this case our method gives the
result that the metallic phase is unstable against any amount of
disorder (see below), consistent with the now accepted
theoretical results \cite{lee,gang4}.

One may wonder why we are in a position to investigate the
Anderson-Hubbard model when there is still not a good
understanding of even the pure system. Our response to this
question is that, as discussed previously
\cite{lizma,lizwa}, and as will be emphasized in the
section to follow, disorder is in fact an advantage for our
investigation.  By sampling a large ensemble of
random configurations of the potential $\{W_i\}$,
the problem of losing long range quantum correlations due to
breaking the system into blocks in the real space RG may be
partially compensated. Also, the disorder averaging allows
us to treat the (average) particle $\overline{n}$ as a
continuous variable---which is not possible in the absence
of disorder.  Finally, even for disordered
systems there already exist some known cases where one
can test the method, as discussed above. These considerations
give us some confidence that our real space RG
approach to the disordered problem is a suitable choice, at least
for our present purpose of investigating the phase diagram.

The rest of the paper is organized as follows: the real space RG
method is described in some detail in the next section;
in Section III, we present our results; and we offer a summary and
discussion in Section IV.

\section{Method}

Our real space RG method is similar to that of Refs.\cite{ma}
and \cite{lizma}.  This real space RG is implemented numerically
on a finite lattice.  The random field $\{ W_{i} \}$ is obtained
through a gaussian random number generator.
Briefly, the RG procedure can be described in
the following five steps: ($i$) Divide the lattice into
blocks of size $n_s$. ($ii$) Compute the block fermion
operators which are defined in terms of four eigenstates
of the block Hamiltonian. Each block is characterized by
an effective on-site potential and an on-site repulsion
between the `block particles'. ($iii$) Determine the
hopping parameters for the block particles from the
inter-block couplings between the site variables.
However, since the block parameters arise from the random
$\{ W_{i} \}$, and so have in general a different distribution
from the original one, we need to ($iv$) repeat the
above procedures [($ii$) and ($iii$)] for a large random
ensemble to determine the distribution of the block parameters. We shall
limit ourselves to tracking only the first two moments
of each distribution (see below).  At this
stage, the Hamiltonian is mapped back onto its original
form with renormalized parameters:
\begin{equation}\label{hahR}
{\widetilde
H}=\sum_{\scriptstyle<\alpha\beta>\atop\scriptstyle\sigma}
   ( {\tilde t}_{\alpha\beta}{\tilde c}^{\dagger}_{\alpha\sigma}
   {\tilde c}_{\beta\sigma} + H.c. )
   + \sum_{\alpha\sigma}({\widetilde W}_{\alpha}-{\tilde \mu})
   {\tilde n}_{\alpha\sigma}
   +{\widetilde U}\sum_{\alpha}{\tilde n}_{\alpha\uparrow}
   {\tilde n}_{\alpha\downarrow}+ {\rm constant}
\end{equation}
where $\alpha$ and $\beta$ are block indices.
Finally, ($v$) we the iterate above sequence
to find the flow and fixed point(s) of the RG.

Now we elaborate each of the above steps:

($i$). The blocks we used in the present work are shown in
Fig.~1. Each of them is chosen to have an odd number of sites,
to allow us to correctly treat the physics of half filling
\cite{oddnum}.  Even for such small blocks the numerical
diagonalization is non-trivial, due to the large Hilbert space
and the necessity of sampling a large number of random
configurations.  For the $3 \times 3$ square lattice at half
filling, for instance, the dimension of the Hilbert space is
$15876\times 15876$. (Here the lattice symmetry is destroyed
by disorder and thus can not be used to reduce the size of the
Hilbert space.) Both types of blocks shown in Fig.~1 have been
tested, and we find that they give the same qualitative physics.
Thus for our present purpose (exploring the phase diagram) we may
focus on the `star' block (Fig.~1a) which is computationally
more convenient.

($ii$). Each (microscopic) site can have one of four possible
states: the no-electron state $|0\rangle$,
the up-spin electron state $|\!\!\uparrow\rangle$,
the down-spin electron state $|\!\!\downarrow\rangle$,
and the two-electron state $|\!\!\uparrow\!\downarrow\rangle$.
The energy for the no-electron state is denoted $E^{(0)}$,
and the two-electron state energy $E^{(2)}$.
The up- and down-spin electron state energies are degenerate and
denoted by $E^{(1)}$. For each block, we find the exact ground
state and ground-state energy
for the Hamiltonian (Eqn. \ref{hah}) for every possible odd
number of particles, restricted to the subspace with
$S=\frac{1}{2}$ (the two $S_{z}=\pm\frac{1}{2}$
states are degenerate). The lowest energy among these ground
state energies then gives us $E^{(1)}$.  Letting $N_{\alpha}$
be the number of particles which gives $E^{(1)}$ in block $\alpha$,
and $|\!\uparrow_\alpha\rangle$ the corresponding ground state,
we then take the $N_{\alpha}-1$ ground state (from the subspace
$S=0$) as $|0_\alpha\rangle$ (with energy $E^{(0)}$), and the
$N_{\alpha}+1$ ground state ($S=0$) as
$|\!\uparrow\!\downarrow_\alpha \rangle$ (with energy $E^{(2)}$).
The block variables may be determined from these states
as follows:
\begin{equation}\label{blkU}
U^{\alpha}=E^{(2)}_{\alpha}+E^{(0)}_{\alpha}-2E^{(1)}_{\alpha}
\end{equation}
\begin{equation}\label{blkW}
W^{\alpha}-\mu^{\alpha} = E^{(1)}_{\alpha}-E^{(0)}_{\alpha}.
\end{equation}

($iii$). The hopping energy between two neighboring blocks
is obtained from the hopping energy of the neighboring sites
on these two blocks.  We calculate the matrix elements in the new
states by insisting that those states are the same for both the
new and the old Hamiltonians. There are four non-zero matrix
elements for $t$ for each bond between blocks $\alpha$ and
$\beta$:
\begin{equation}
{\tilde t_{\alpha\beta}^{\scriptscriptstyle (1)}}=
       \langle 0_{\alpha}\uparrow_{\beta}
            |\,{\widetilde H_t}\,|
       \uparrow_{\alpha}0_{\beta}\rangle=
       \langle 0_{\alpha}\uparrow_{\beta}
            |\,H_t\,|
       \uparrow_{\alpha}0_{\beta}\rangle
\end{equation}
\begin{equation}
{\tilde t_{\alpha\beta}^{\scriptscriptstyle (2)}}=
       \langle \uparrow_{\alpha}\,\downarrow_{\beta}
            |\,{\widetilde H_t}\,|
       \,0_{\alpha}\uparrow\!\downarrow_{\beta}\rangle=
       \langle \uparrow_{\alpha}\,\downarrow_{\beta}
            |\, H_t\,|
       \,0_{\alpha}\uparrow\!\downarrow_{\beta}\rangle
\end{equation}
\begin{equation}
{\tilde t_{\alpha\beta}^{\scriptscriptstyle (3)}}=
       \langle\uparrow\!\downarrow_{\alpha}0_{\beta}\,
            |\,{\widetilde H_t}\,|
       \downarrow_{\alpha}\,\uparrow_{\beta}\rangle=
       \langle\uparrow\!\downarrow_{\alpha}0_{\beta}\,
            |\, H_t\,|
       \downarrow_{\alpha}\,\uparrow_{\beta}\rangle
\end{equation}
\begin{equation}
{\tilde t_{\alpha\beta}^{\scriptscriptstyle (4)}}=
       \langle\uparrow\!\downarrow_{\alpha}\downarrow_{\beta}\!
            |\,{\widetilde H_t}\,|
       \downarrow_{\alpha}\,\uparrow\!\downarrow_{\beta}\rangle=
       \langle\uparrow\!\downarrow_{\alpha}\downarrow_{\beta}\!
            |\, H_t\,|
       \downarrow_{\alpha}\,\uparrow\!\downarrow_{\beta}\rangle.
\end{equation}
There are 3 bonds allowing hopping between any two star blocks
$\alpha$ and $\beta$ (also in the square blocks; see Fig.~1).
We sum these three hops to get ${\tilde t_{\alpha\beta}^{(i)}}$.
The above 4 ${\tilde t_{\alpha\beta}}$'s are then averaged,
$ \tilde t_b = {\frac{1}{4}\sum_{i=1}^4{\tilde
t_{\alpha\beta}^{(i)}}}$, for one connection $ \tilde t_b$
between two blocks $\alpha$ and $\beta$.

($iv$). Now we average over an ensemble of random configurations
to determine the distribution of the parameters in the block
Hamiltonian. Since (\ref{blkU}) is always positive (as verified
numerically), we simply use its mean (denoted as $\widetilde U$)
as the renormalized on-site repulsion between the block particles.
The renormalized chemical potential $\tilde \mu$ is defined
by the mean of the right hand side of (\ref{blkW}). This
implies that the renormalized random potential
$\{W^\alpha \}$ has zero mean. However,
the distribution of the block potential is in general
different from the original one. Here we choose to keep
track of only the first two moments, the mean ($\equiv 0$)
and the variance ($\equiv {\widetilde W}$); we thus
map the renormalized distribution of the $\{W_i\}$
back onto an (independent) gaussian form.

The determination of
the block hopping parameter is more subtle. Consider a
block (of any size) without disorder. In general, the
ground state is degenerate, where the degeneracy is related
to the symmetry of the lattice and to the fermi statistics of the
electrons. Any amount of disorder breaks this lattice
symmetry and therefore lifts the degeneracy. As (degenerate)
perturbation theory shows, depending on the
configuration of the random fields $\{W_{i}\}$, the sign
of the ${\tilde t_{\alpha\beta}}$'s can be either positive
or negative. This causes {\it frustration\/} when the product
of ${\tilde t}_{\alpha \beta}$'s around a closed path is
negative. This is different from the boson case considered in
Ref.\cite{lizma} where the ground state of the pure system
is non-degenerate and the kinetic energy is unfrustrated
by site-disorder. To take this effect of frustration into
account, one has thus to keep track of the lattice structure.
More specifically, for the `star' block considered here,
we first build a square lattice
which consists of $n_{b} = 125$ coupled star-blocks
($n_{s} \times n_{b} =625$ sites).
Step ($iii$) is then performed to obtain the corresponding
block hopping parameters. This explicit lattice structure
enables us to compute the frustration index
(defined by the ratio of number of frustrated plaquettes
to the total number of plaquettes).  It turns out that,
starting from a uniform hopping constant $t_{ij} =t$,
the RG described above will randomize the hopping parameter
and frustrate the kinetic energy. Regardless of the starting
configuration, the frustration index for the block system
is always near 0.5. We approximate the block hopping
parameters with an independent
gaussian distribution which is determined by their mean
$\tilde t_{ave}$ and variance $\tilde t_{var}$.
In the actual calculation, we typically average over 5--10
such lattices. Due to the symmetry of the square lattice,
one can always choose the mean of the hopping parameter
$\tilde t_{ave}$ to be positive. Under the RG iterations for the
2D problem, $\tilde t_{ave}$ decreases rapidly, reflecting the
frustration.

($v$). Using the new set of parameters one can repeat the
process described above and study the flow of the parameters under
the RG iteration.
Physical phases are identified from the stable fixed points
of the RG.

The Hamiltonian (\ref{hahR}) is characterized by four
independent parameters, which may be chosen as
${\tilde t}_{ave}/\widetilde{W},\widetilde{U}/\widetilde{W},
\tilde{\mu}/\widetilde{W}$, and ${\tilde t}_{var}/\widetilde{W}$.
Among the others, ${\tilde t}_{var}/\widetilde{W}$ is found
to always renormalize to zero; hence we shall not discuss
this parameter further in the following.
For simplicity we label the relevant dimensionless parameters
as follows: $\widetilde{U}/\widetilde{W} \equiv U/W $ gives the
strength of the repulsion; the flow of $\tilde
t_{ave}/\widetilde{W} \equiv t/W$
indicates insulating ($\to 0$) or metallic ($\to\infty$)
behavior; and $\tilde{\mu}/\widetilde{W} \equiv \mu/W$ gives a
dimensionless measure of the chemical potential.

The choice of the block states described in ($ii$) needs some
more explanation. We truncate the Hilbert space of the block
by choosing four low lying states such that the block Hamiltonian
and the site Hamiltonian have the same form. However, there are
different possibilities of choosing these 4 states.
The simplest possibility (implemented in Ref.~\cite{ma} at half
filling) is to always pick the same $N_{\alpha}=N$ for every block
$\alpha$.  This ``fixed-$n$'' (where $n$ is defined by
$n \equiv N_{\alpha}/n_{s}$) procedure is the most artificial
of the procedures we have used; however, one might argue that
it is adequate for incompressible states.

A second (``fixed-$\mu$'') procedure is to fix the chemical
potential $\mu$ rather than the density $n$.
This procedure can only work however if the chosen value for $\mu$
corresponds to a density which can be represented on the blocks
by an integer particle number; otherwise, instabilities occur
in the RG flow \cite{lizma}. Hence we have only used the
fixed-$\mu$ procedure at half filling. In this case, the chemical
potential is known to be precisely $U/2$ and can therefore be set at
the beginning of the RG iteration. Since the distribution of the
random potential $\{W_{i}\}$ is symmetric with zero mean, the
statistical fluctuations will then preserve the average density
at $1$ and $\tilde \mu$ at ${\widetilde U}/2$ by averaging
over the random configurations. Since the density is allowed to vary
from block to block, this method allows one to study compressible as
well as incompressible states.

The possibility of allowing the density to fluctuate
in such a real space RG scheme is unique to disordered
systems. In a pure system, all blocks are identical and
$N_{\alpha}$ will thus be chosen the
same for all blocks. The particle number fluctuation in any
given region (of any size) is thus one, so that the RG
can only describe an incompressible state \cite{lizma}.
Thus one expects that such a real space RG scheme for
pure systems will be mostly applicable at half filling,
where it is indeed incompressible \cite{pfeutyPURE}.

We have used a third procedure to study
the physics in a region around half filling, by
allowing the chemical potential to also flow in the RG iterations.
This enables us to explore the RG flow in the full 3D parameter
space $(t/W,U/W,\mu/W)$.  In the absence of disorder, this
would not be possible, since the density cannot vary in any
block. In the presence of disorder, one selects $N_{\alpha}$
which minimizes the energy in a given block $\alpha$, and thus allows
$\overline n $ (where again the overbar mean disorder average)
to vary continuously along with $\mu$. However this flow must
fail at high and low densities, where the small size of our blocks
imposes strict upper and lower bounds on the range of densities
which can be handled by the method---for example, the density in
a star block cannot fall below 0.2 nor exceed 1.8.
Hence with this procedure it is necessary to follow
the flow of $\overline n $ as well, and to discard flows
when $\overline n $ saturates at its upper or lower bound.
With this procedure it is sometimes convenient to parameterize
the chemical potential as $\tilde{\mu}/\widetilde{U} \equiv
\mu/U$ since it is in terms of this parameter that we can locate
half filling ($\overline n = 1$ at $\mu/U = 1/2$).

\section{Results}

We start by considering the Anderson model ($U=0$).
In this case, with either the fixed $n$ or the fixed
$\mu$ method, the only meaningful parameter of the system
is $t/W$.  Either procedure gives the same qualitative result:
we find two fixed points, at $t/W=0$ and at $t/W = \infty$.
The fixed point describing the pure system ($t/W=\infty$)
is unstable, with flow towards the attractive
(insulating, $t/W=0$) fixed point.  Thus we find that,
for noninteracting fermions, disorder is always relevant,
in agreement with the prediction from scaling theory
\cite{gang4,wegner}.

Next we consider the interacting case $U\ne 0$.
In Fig.~2, we show RG flow diagrams in the two-dimensional
parameter space $(t/W,U/W)$.  Fig.~2a is obtained at
half filling. Here we again find that the two RG procedures
(fixed-$n$ and fixed-$\mu$) give similar results.
Fig.~2b depicts RG flow at densities away from half filling,
obtained using the fixed-$n$ procedure.
The possible densities are $n=1/5,3/5$ for the star blocks, and
$n=1/9,3/9,5/9,7/9$ for the $3\times 3$ square blocks
(apart from those which may be obtained using particle-hole
symmetry). They all give qualitatively the same flow diagrams.
At half filling, Fig.~2a shows that, apart from the unstable
fixed points describing the noninteracting ($U=0$) and pure
($t/W=\infty$) phases,
there are two stable fixed points at ($t/W=0,U/W= \infty$) and
at ($t/W=0, U/W \approx 1.3$). Between these two phases is a
separatrix which terminates at a repulsive fixed point
($t/W=0$, $U/W= (U/W)^* \approx 7.3$).  Away from half
filling and at fixed density, the RG flow has only one stable
fixed point, at finite $U/W$. We note that, with respect to the
noninteracting system, $U$ is relevant at all the fillings we
examined.

We next ask, what is the nature of the various phases revealed
by the stable fixed points in Fig.~2?
Since $t/W$ renormalizes to zero in all the cases, there is no
metallic state. However, the nature of the insulating states
needs some elaboration. We consider
first the half-filled case. For the pure system (Hubbard
model) at half filling, it is believed that the system is
always insulating with long-ranged antiferromagnetic order
for any finite $U>0$.
However, while the disordered system is always insulating, the
physics responsible for the insulating behavior may vary.
This situation is best illustrated in the essentially exact
calculations on the infinite dimensional Hubbard model
\cite{mark}: at low but above the Neel temperature, there
is a critical value of $U=U_c$, beyond which a gap opens
up in the quasi-particle spectrum and the system changes
from a metal to a Mott-insulating state. Such a paramagnetic
solution persists down to $T=0$, although it becomes unstable
at low temperature and the true ground state is
antiferromagnetic for any finite value of $U$. Thus one may
expect that, while for small $U$ the insulating state
is a result of the delicate (antiferromagnetic)
correlations, at large $U$ it is simply
due to the large energy cost for double occupancy.
Upon introducing disorder, the Mott transition
(masked by the antiferromagnetic long range order in the
pure case) is revealed, but the corresponding metallic state
now becomes insulating also.
Hence we interpret the state described by the fixed point
at $(U/W) \rightarrow \infty$ as the Mott-Hubbard phase,
while that associated with the fixed point at finite $U/W$
we will call the `Anderson' phase, since it is expected (to be
corroborated below) to be gapless.

Remarkably, this phase diagram is quite similar to that
for the 1D case, which was calculated at half filling in
Ref.\cite{ma} using the fixed-$n$ method. (We have obtained
the same picture for the 1D Anderson-Hubbard problem using the
fixed-$\mu$ method at half filling.)
The value of the unstable fixed point separating
the two insulators is about
$(t/W,U/W)^*\cong(0,7.3)$. This value for the 2D system is very
close to the fixed point obtained by Ma for the 1D case
[$(t/W,U/W)=(0,8.3)$], and to the slope of critical line for the
opening of a compressibility gap in the ($U_c,W$)
plane obtained by Dom\'\i nguez and Wiecko (DW)
\cite{dw} for the 3D case, $U_c\cong6.7W(W/t\rightarrow\infty)$.
As mentioned before, in getting the hopping parameter
$\tilde{t_b}$ between two blocks, we take an arithmetic
average over 4 $\tilde{t}_{\alpha\beta}$'s
to take account of frustration. The
$\tilde{t}_{\alpha\beta}$'s tend to be of varying sign
(for reasons discussed earlier) and hence to
cancel each other when we average, so that the parameter
$t/W$ rapidly approaches zero as the RG iteration proceeds. The
slope of the separatrix between the two insulators is,
therefore, nearly zero (unlike the 1D case \cite{ma}).

The stable fixed point found (with fixed density) away from
half filling can be interpreted naturally as a fixed point
describing the Anderson insulating phase. Although this
phase diagram is obtained using the RG procedure for
fixed $n$---which is more appropriate for incompressible
states---we believe that this Anderson phase is
actually compressible from the physical point of view.
The compressibility of this phase at or near half
filling may be investigated through the RG flow in the 3D
parameter space $(t/W, U/W, \mu/W)$.

We have studied the compressibility in our numerical
RG calculations using two methods. One, which is purely
heuristic, is to stop the calculation after a single
iteration, associating the renormalized values of density
and chemical potential with the (fixed) values of $U$, $t$,
and $W$ which were input. This method---which can not probe
the long wavelength physics seen from repeated RG
iterations---nevertheless gives surprisingly good results,
possibly due to the large size (625 sites) of the finite lattice
which we used, coupled with the further averaging over disorder.
Results obtained using this method are plotted in Fig.~3.
We see that (thanks to the disorder and the averaging) the density
$\overline n$ flows smoothly with the chemical potential, with two
significant exceptions. One exception occurs when the density {\it
saturates\/} at the maximum or minimum value allowed by the finite
size of our block (i.e., for the star block, 0.2 and 1.8). This
saturation marks a limit beyond which our method gives meaningless
results. The other departure from smoothness occurs at
$\overline n = 1$, for sufficiently large $U$, and is due to
the abovementioned incompressibility.  The incompressibility
(the Hubbard gap) is broadened with increasing Coulomb repulsion
(Fig.~3). In the inset we plot the Hubbard gap $\Delta\mu/t$ as a
function of $U/t$, with fixed $t$ and $W$. We can see that the
Hubbard gap increases linearly with $U$. The slope $\alpha$
[$\equiv \frac{d(\Delta\mu)}{dU}\Bigr|_t$] of the three curves
is about $\alpha\sim 1.0$, which is consistent with what is
expected, and also with the result obtained by DW \cite{dw}.

 From the inset of Fig.~3 we can write $\Delta\mu = \alpha(U-U_c)$
at fixed $W$.  Then in ($\mu/U$,$U/W$) parameter space the gap can
be described (for fixed $t$) by
\begin{equation}\label{gapoverU}
\frac{\Delta\mu}{U}=\alpha(1-\frac{U_c}{U})\;\;\;
\mbox{for}\;\;\; (U\geq U_0).
\end{equation}
Taking the large $U$ limit, the gap $\Delta\mu/U$ will approach
$\alpha$ in $\mu/U$ and $U/W$ space.
Hence our second method for studying the behavior of the
incompressibility is to follow the RG flow in the 3D parameter space
$( t/W, U/W, \mu/U )$, distinguishing however those flows which remain
pinned at $\overline n = 1$ from those which do not.
Fig.~4 is a 3D flow diagram inside the limits at which the
density saturates, but projected out onto the 2D plane $(U/W,\mu/U)$.
(We project out the behavior of $t/W$ since it is the most predictable,
always flowing to zero.) In this figure we see that the
Mott-Hubbard phase (shown by the shaded area) with
$\overline n = 1$ is bounded by equation (\ref{gapoverU}).
We can also see the two fixed points in the plot, which correspond
to those in Fig.~2a. For $U/W\to\infty$ we have a ``fixed bar''
(i.e., a {\it line\/} of fixed points in $\mu$--$t$--$U$ space)
which attracts the flows inside the Mott-Hubbard phase. (This fixed
bar is of course a fixed {\it point\/} in $n$--$t$--$U$ space,
with $n=1$.) In the Anderson phase we can see that the
parameter $\mu/U$ flows to the value for half filling $\mu/U=1/2$.
Hence we find that there is a finite region in density, around
$\overline n = 1$, in which the system is a compressible
(Anderson) insulating phase and is characterized by the fixed
point at half filling.

Since the discovery of high $T_c$ superconductivity,
there have been suggestions that the dimensionality alone
will invalidate the Fermi liquid theory and make a 2D
interacting system a highly correlated one \cite{anderson3},
as it does for its 1D counterpart. While for the 2D Hubbard
model near half filling there is little doubt that the system
is indeed highly correlated, controlled perturbative expansions
suggest that at low density a system of interacting 2D fermions
can be well described by the conventional Fermi-liquid
theory \cite{FL2D}. If this is the case, the insulating
state at low density would be simply due to the localization
of the quasi-particles. From this point of view, one would
expect that the (compressible) insulating state near half
filling will be quite different from that at low (or high)
densities, so that there should be an additional stable fixed
point describing such a `conventional' Anderson insulating
state, distinct from the `highly correlated' Anderson insulating
state controlled by the fixed point at half filling.
Hereafter we shall call the `conventional' Anderson phase
an `Anderson-Fermi' insulator; the (presumably) highly
correlated Anderson phase described by the Anderson fixed
point at half filling we call the `Anderson-Luttinger' insulator.
We choose the latter name
since our RG study shows a strong resemblance between the 1D and
2D systems near half filling, and since generic (pure) 1D
systems are described by the highly correlated
`Luttinger liquid' \cite{haldane}.  In our studies we have
found no evidence for an Anderson-Fermi insulating phase
characterized by a high- or low-density fixed point. We note
however that we cannot rule out the existence of
such fixed points, since our method is only reliable in a
finite range of densities around half filling, and so may be
incapable of detecting these uncorrelated phases.

In Fig.~5 and Fig.~6 we show the full flow diagrams in the 3D
parameter space. Taking account of particle-hole symmetry, we omit the
region $\overline n < 1$. We also omit $U<0$ (which is expected to
give different physics) and $t<0$ which is trivially related to $t>0$;
hence we are left with one octant of the full space.
In each plane the flows are the projection of 3D parameter flows.
The incompressible Mott-Hubbard phase is lightly shaded
in the $(\mu/U,U/W)$ plane.
Our RG approach fails in a high- and a low-density region; the former
is shown in Figs.~5 and 6 with dark shading. In this dark region we
can imagine either two fixed points (one in the high-density region
and the other one in the low-density area), or
none. In the case of no fixed points in the dark area (Fig.~5)---or if
the noninteracting fixed points are unstable to any finite $U$---the
RG parameters outside the Mott-Hubbard phase flow to the one fixed
point, so that there is only one (``correlated'') Anderson phase
in the Anderson-Hubbard model.
On the other hand, if there are two stable fixed points (at high and
low density) (Fig.~6), the Anderson-Hubbard problem will have two
different Anderson phases as discussed above.  We include both
Figs.~5 and 6 because we do not believe that we can distinguish
these two scenarios within the limitations of our method.

\section{Summary and Discussion}

Using a quantum real space renormalization group method,
we have obtained the phase diagram of the 2D Anderson-Hubbard
model near half filling at $T=0$.
A test of our method for noninteracting fermions shows the
instability of the metallic phase for any nonzero disorder $W$,
in agreement with commonly accepted results and hence
providing some evidence that our method is qualitatively
reliable. By studying the renormalization of chemical potential
and the corresponding flow of particle density,
we were able to estimate the compressibility gap $\Delta\mu$
in the Mott-Hubbard phase as a function of $U$ and $W$.  We
found that $\Delta\mu$ increases proportional to the interaction
$U$ (at constant $W$) above a critical value ${(U/t)}_c$, with
the constant of proportionality $\alpha$ about 1, and that
$\Delta\mu$ decreases with increasing $W$ (at fixed $U$).  Our
results here are also in good agreement with those obtained by
other methods, which gives us further confidence in our approach.

Our studies show that there is no metallic phase for the 2D
Anderson-Hubbard model, for any finite value of the random
potential $W$ and of the repulsive interaction $U$ between
the electrons.  The interplay between interaction and
disorder yields two insulating phase at half filling:
an incompressible Mott-Hubbard insulator and a gapless
Anderson insulator.  The phase diagram strongly resembles
that for the corresponding 1D system.
Away from half filling, the insulating phase is always
gapless, and its properties are controlled by the fixed
point describing the Anderson insulator at half filling.
We characterize such a highly correlated insulating phase
as the `Anderson-Luttinger' insulator. $U$ is relevant with
respect to the noninteracting fixed point for all the cases
we have considered.  We would like to emphasize, however,
that the relevance of $U$ itself does not constitute
evidence for the existence of the `Anderson-Luttinger'
insulator.  Since the noninteracting disordered
system is described by the localized ($t/W = 0$) fixed point,
one may expect that, at least in the low doping (near half
filling) case, any interaction will be relevant regardless
of the properties of the corresponding pure system, although
there is a possibility that short-ranged interactions such
as the on-site $U$ studied here are not relevant in the
dilute (low or high density) limit.  The dilute fixed point,
which presumably describes the conventional `Anderson-Fermi'
insulator whose physics is related to that of localized
noninteracting fermions has not been found within our
approach, which is however limited to a range of
densities around half filling.

The picture that we find at half filling is not unexpected: the
instability of the noninteracting fixed point, the consequent
finite-$U$ (gapless) fixed point, and the opening up of a gap
at $U_c$ with flow towards $U=\infty$ in the Mott-Hubbard phase.
The one feature of our results that is perhaps somewhat unexpected
is the existence of a finite region in $n$ (or $\mu$) around
half filling, which is dominated by the $n=1$ fixed point.
One could imagine a different result, namely, that, like the
Mott-Hubbard phase, the `Anderson-Luttinger' phase is well-defined
only at or close to half filling, becoming unstable as $n$
deviates from this region and flowing towards a `dilute' fixed
point. In other words, one could imagine that
$\delta \mu \equiv \mu - U/2$ is relevant when its magnitude
becomes nonzero or sufficiently large, which is not what we
have found for the region allowed by our RG scheme.

It is conceivable that this result might be due to an artifact
of our method. That is, one might conjecture that the flow
towards half-filling reflects only the stability of the
{\it algorithm\/} at half filling, rather than the stability
of the thermodynamic phase.  Although we see no reason for
this to be the case, we cannot rule out this possibility.
We do however gain some confidence in the results of our RG method,
in the case where we allow the chemical potential to flow, from the
good agreement of these results for the Hubbard phase with
existing results obtained by other methods.  We note that these
results (Figs.~3 and 4) were all obtained using this algorithm.

We therefore assume that the correlated `Anderson-Luttinger'
insulator indicated by our results is in fact the true ground
state of the 2D Anderson-Hubbard problem for some region around
half filling. This suggests a number of directions for future
work. It is clearly important to try to clarify the nature of this
phase, in both the 2D and the 1D problems, for instance by calculating
density-density or magnetic correlation functions and their RG flow.
Furthermore, if indeed the physics for this disordered problem
around half filling is described by the Anderson fixed point at
half filling, one can expect to gain significant information
about the lightly doped case by directly studying the half-filled
case (where, for instance, there is no ``sign problem'' in quantum
Monte Carlo simulations).
It would also be of considerable interest to extend this
work to the 3D problem, where one expects a metallic phase, and
metal-insulator transitions of various types \cite{ma}.
Unfortunately, the smallest isotropic 3D block with an odd site number
($3\times3\times3$) \cite{oddnum} is far too large for exact
diagonalization.  Since the 3-(spatial)-D problem is of interest
both in its own right, and as a further test of the present 2D
results, we believe that the problem of extending our real space
RG technique for disordered systems to the 3D case merits some
further effort.

\noindent{\bf Acknowledgements.} We thank M. Ma and M. Randeria
for useful discussions and comments.  This research was supported
by the NSF under Grant \# DMR-9101542, and by the U.S.
Department of Energy through Contract No. DE-AC05-84OR21400
administered by Martin Marietta Energy Systems Inc. Part of the
computations were carried out on the MASPAR MP-2 computer located
at the University of Tennessee; we thank Christian Halloy of the
Joint Institute for Computational Science (JICS) for help in
the use of the MP-2.

\begin{figure}
\caption{Two types of blocks used in our RG approach for the square
lattice. (a) star lattice (b) $3\times 3$ lattice.}
\label{fig1}
\end{figure}

\begin{figure}
\caption{Flow diagrams of the 2D Anderson-Hubbard model as obtained by
our RG approach, at fixed filling of the lattice.
(a) For half filling  we see two insulating phases: a Mott-Hubbard
phase at large $U/W$ and an Anderson (gapless) phase at smaller $U$.
(b) Away from half filling we see only the Anderson phase.}
\label{fig2}
\end{figure}

\begin{figure}
\caption{Density versus $\mu/U$ with different $U/W$ for fixed
$t(=0.5)$ and $W(=1.0)$. Each curve is shifted to the right, for
visual purposes, by $0.5$ as $U/W$ is increased by $1$. Density is
pinned at $\overline n=1$ over a finite range in $\mu$ as $U$ is
increased. (inset) The incompressibility $\Delta \mu$ in the
Mott-Hubbard phase is plotted as a function of $U$ in units of $t$,
for fixed $W$ and $t$. For comparison, we include some results for the
1D lattice.}
\label{fig3}
\end{figure}

\begin{figure}
\caption{ Projection on a plane of flows of 3D RG parameters
$\{t/W,U/W,\mu/U\}$. In our projection the flow of $t/W$ is not shown,
since this parameter always flows to zero.
The Mott-Hubbard phase (in which the density is pinned at 1)
is shaded. The unstable fixed point ($U/W = 7.3$, $\mu/U=0.5$)
marking the boundary of the Mott-Hubbard phase
is marked with a small circle; the stable one is located at
$(U/W,\mu/U)=(1.3,0.5).$ }
\label{fig4}
\end{figure}

\begin{figure}
\caption{3D parameter flow diagram. In the dark area, where the
density is saturated and so our method gives no information,
we are assuming the chemical potential flows toward
half filling. The Mott-Hubbard phase is marked with light shading.}
\label{fig5}
\end{figure}

\begin{figure}
\caption{3D parameter flow diagram, but with an alternative
hypothetical scenario from that shown in Fig.~5. Here we assume the
existence of attractive fixed points in the dark
areas (high and low density), indicating the presence of
``uncorrelated'' insulating phases distinct from the ``correlated''
phase we find around half filling. Since our method fails in the dark
shaded area, it cannot distinguish between the picture shown here and
that in Fig.~5.}
\label{fig6}
\end{figure}


\begin{references}
\bibitem{oddnum}
Blocks with an even site number $n_s$ can also be employed
with our method. However, since there is no odd $N$ such that
$n \equiv N/n_{s} =1$ in the corresponding pure system, one would
not find the Mott-Hubbard phase in this case.
(The importance of odd $N$ in our method arises from the fermi
statistics, as explained in step ($ii$) in the text.)
\end{references}
\end{document}